\documentclass[runningheads]{llncs}
\usepackage[T1]{fontenc}
\usepackage{orcidlink}
\usepackage{helvet}
\usepackage{courier}
\usepackage{graphicx}
\usepackage{float}
\usepackage{algorithm}
\usepackage{algpseudocode}
\usepackage{enumitem}
\usepackage{tikz,amsmath}
\usepackage{hyperref}
\usepackage{amsmath,amssymb}
\usepackage[table]{xcolor}

\usepackage{url}

\usepackage{pgfplots}
\pgfplotsset{compat=1.18}

\usepackage{caption}
\DeclareMathOperator*{\argmin}{arg\,min}

\usepackage{booktabs}      
\usepackage{multirow}      
\usepackage{rotating}      
\usepackage{graphicx}      
\usepackage{array}         
\begin{document}
\title{FedSurrogate: Backdoor Defense in Federated Learning via Layer Criticality and Surrogate Replacement}
\titlerunning{FedSurrogate}
%
\author{Fatima Z. Abacha \inst{1}\orcidlink{0000-0001-5285-5113} \and
Sin G. Teo \inst{2}\orcidlink{0000-0003-1090-505X}\and
Yuanxiang Wu \inst{3}\orcidlink{0009-0007-6183-6724}\and
Lucas C. Cordeiro \inst{1}\orcidlink{0000-0002-6235-4272}  \and 
Mustafa A. Mustafa \inst{1,4}\orcidlink{0000-0002-8772-8023} 
}

\authorrunning{F. Abacha et al.}
%
\institute{The University of Manchester, UK \and
Institute for Infocomm Research, A*STAR, Singapore
 \and
 Monash University, Australia
 \and
COSIC, KU Leuven, Belgium \\
\email{{fatima.abacha}@postgrad.manchester.ac.uk}}
%
\maketitle              
%

\begin{abstract}
Federated Learning remains highly susceptible to backdoor attacks--malicious clients inject targeted behaviours into the global model. Existing defenses suffer from substantial false-positive rates under realistic non-independent and identically distributed (non-IID) data, incorrectly flagging benign clients and degrading model accuracy even when adversaries are correctly identified. We present FedSurrogate, a novel backdoor defense that addresses this limitation by combining bidirectional gradient alignment filtering with layer-adaptive anomaly detection. FedSurrogate performs selective clustering on security-critical layers identified via directional divergence analysis, concentrating the detection signal on a low-dimensional subspace. A bidirectional soft-filtering stage screens trusted clients for residual contamination while rescuing false positives from suspects, substantially reducing misclassifications under heterogeneous conditions. Rather than removing confirmed malicious updates, FedSurrogate replaces them with downscaled surrogate updates from structurally similar benign clients, preserving gradient diversity while neutralising adversarial influence. Extensive evaluations demonstrate that FedSurrogate maintains false-positive rates below 10\% across all datasets and attack types, compared to 31-32\% for the nearest comparably effective baseline, while achieving superior main-task accuracy and maintaining attack success rates below 2.1\% across all tested datasets and attack types under challenging non-IID settings.

\end{abstract}


\section{Introduction}

\label{sec:intro}

Federated learning (FL)~\cite{mcmahan2017communication} enables multiple parties to collaboratively train a shared model while keeping their data local, making it appealing for privacy-sensitive domains. 
However, the same client autonomy that preserves privacy also allows adversaries to poison their local updates, thereby compromising the global model~\cite{tolpegin2020data}. 
A particularly insidious variant is the {backdoor attack}, in which an adversary embeds a hidden trigger during training so that the global model behaves normally on clean inputs but consistently misclassifies any trigger-bearing input into an adversary-chosen label~\cite{bagdasaryan2020backdoor,sun2019can}. Because such triggers remain dormant under benign conditions and readily evade anomaly-based defenses, backdoor attacks pose a key threat to the trustworthiness of FL systems~\cite{naseri2024badvfl}.

Several approaches have already been proposed to counter backdoor attacks. Cosine-similarity methods (FoolsGold~\cite{fung2018mitigating}) 
assume that malicious updates are highly correlated, 
which does not hold for non-independent and identically distributed (non-IID) data.
Trust-based approaches (FLTrust~\cite{cao2020fltrust}) 
require clean server-side reference data, which is often unavailable in practice. More recent defenses (FLAME~\cite{nguyen2022flame}, FedGrad~\cite{nguyen2023fedgrad}, Snowball~\cite{qin2024resisting}, AlignIns~\cite{xu2025detecting}, FLShield~\cite{kabir2024flshield}) 
integrate techniques such as clustering, clipping, noise injection, spectral analysis, and sub-model validation to improve robustness. While these methods demonstrate empirical viability, they remain sensitive to statistical heterogeneity across clients and often incur high false-positive rates (FPRs), leading to the exclusion of benign participants and a substantial decline in global model performance, while alignment-based variants face a complementary weakness against stealth-optimised attacks (Neurotoxin~\cite{zhang2022neurotoxin}) that embed the backdoor in low-magnitude coordinates while preserving the overall gradient direction. Moreover, existing defenses typically treat all model layers uniformly during detection, ignoring the well-documented observation that backdoor attacks concentrate their poison in a small subset of layers~\cite{zhuang2024backdoor}, thereby diluting the adversarial signal across millions of clean parameters and reducing detection sensitivity. Recent reweighting-based aggregators (SPMC~\cite{he2025spmc}) attempt to recover utility by softening the exclusion of suspicious clients, but apply their adjustment uniformly across the full parameter vector, conflating the layers that carry the backdoor with those that carry the legitimate task knowledge. In summary, the effectiveness of existing defenses is limited by high false-positive rates, vulnerability to stealth-optimized attacks, and fixed or global views of client updates, which fail to isolate the layers where adversarial signals concentrate.

We propose \textbf{FedSurrogate}, a novel defense that addresses these limitations through three complementary stages: (i)~\emph{Layer Criticality Analysis (LCA)} that measures per-layer directional divergence across client updates to identify the small subset of security-critical layers where backdoor signals concentrate, followed by density-based clustering restricted to those layers; (ii)~\emph{bidirectional soft-filtering} that applies gradient-alignment metrics to 
screen the trusted set for missed adversaries and rescue false positives from the suspect pool; and (iii)~\emph{surrogate-replacement} that substitutes only the LCA-identified critical layers of each confirmed malicious update with the corresponding layers from a structurally similar benign client, preserving the main-task knowledge in the remaining layers and downscaling the resulting surrogate to prevent aggregation imbalance. The design builds on the observation that model parameters faithfully encode the statistical properties of the data on which they are trained, allowing benign client weights to act as surrogates for the data of compromised peers without explicit data sharing~\cite{li2020federated,ghosh2020efficient,fraboni2021clustered}.

Our main \emph{contributions} are: (1) We introduce \emph{Layer Criticality Analysis}, a per-round, directional-divergence-based mechanism for dynamically identifying the layers most informative for distinguishing benign from malicious updates. Unlike prior work that pre-selects layers heuristically or operates on the full parameter vector, LCA localises detection to a low-dimensional, high-signal subspace. (2)  We propose \textit{FedSurrogate}, a three-stage server-side defense built on LCA that combines bidirectional gradient-alignment filtering with surrogate replacement, distinguishing the latter empirically from reweighting-style aggregation. (3) We evaluate \textit{FedSurrogate} across four benchmarks and two architectures against seven defenses under centralised, distributed, and stealth-optimised attacks.
\textit{FedSurrogate} maintains high main task accuracy within 3 percentage points of FedAvg, and attack success rate below 2.1\% across all settings, achieves the lowest FPR on every dataset-attack combination, and degrades gracefully under increased malicious client ratios and larger client populations.

\section{Background and Related Work}
\label{sec:background}

\noindent \textbf{Backdoor attacks in FL.} 
There are different types of such attacks in FL.
\textit{Centralized backdoor attacks (CBA):} a single malicious client modifies a subset of its local samples, e.g., altering pixel regions and relabelling them to a target class, then optimizing its model to fit both benign and triggered data~\cite{bagdasaryan2020backdoor}.
Once aggregated, the global model inherits the trigger-to-target mapping while preserving clean accuracy.
\textit{Distributed backdoor attacks (DBA):}
To enhance stealth and persistence, adversaries decompose the trigger across multiple compromised clients~\cite{xie2019dba}.
Each malicious client trains on a distinct fragment, e.g., a different pixel segment, so that no individual update reveals the full pattern; aggregation reconstructs the complete trigger behaviour in the global model.
This decomposition resists single-client inspection and allows the backdoor to persist even when only a subset of adversaries participates in a given round.
\textit{Neurotoxin:}
Unlike CBA and DBA, which modify inputs directly, Neurotoxin~\cite{zhang2022neurotoxin} targets the \emph{durability} of the backdoor by projecting malicious updates onto parameter coordinates that benign clients rarely update.
By perturbing only the bottom-$k\%$ of gradient magnitudes observed in recent rounds, the adversary avoids being overwritten by subsequent benign aggregation, allowing the backdoor to survive long after the adversary stops participating.
This makes Neurotoxin stealthier and more resilient than attacks that perturb the full parameter space, and poses a harder challenge for defenses that rely on detecting anomalous update directions or norms.

\vspace{0.2em}

\noindent \textbf{Defenses against backdoor attacks in FL.}
\label{subsec:defenses}

Early defenses such as Krum and MultiKrum~\cite{blanchard2017machine} mainly focused on \textit{robust aggregation algorithms} that select the client update(s) closest to the majority in parameter space. Their aim is to make aggregation itself inherently resilient to adversarial contribution. However, their guarantees assume bounded adversary counts and collapse when benign variance, typical under heterogeneous data, mimics adversarial noise. More recent schemes focus on \textit{detection-and-mitigation defenses} in order to 
identify and exclude or down-weight poisoned updates. 

{FoolsGold}~\cite{fung2018mitigating} penalizes clients whose updates exhibit high pairwise cosine similarity, assuming that colluding adversaries produce near-identical directions; it performs poorly under non-IID conditions, where benign updates may also align.
{FedGrad}~\cite{nguyen2023fedgrad} uses final-layer gradient similarity to flag anomalies, but suffers from high FPRs that erode global accuracy.
{FLAME}~\cite{nguyen2022flame} combines HDBSCAN clustering, gradient clipping, and noise injection, but frequently misclassifies benign clients in heterogeneous settings, and its injected noise degrades main-task accuracy.
{FLShield}~\cite{kabir2024flshield} validates client updates against held-out data at trusted participants and filters those whose behavioural signatures deviate from the consensus; its reliance on validation proxies makes it sensitive to benign data heterogeneity, where legitimate updates can appear inconsistent with the reference behaviour.
{Snowball}~\cite{qin2024resisting} couples K-Means clustering with a VAE-based anomaly detector; it performs well on simple datasets (MNIST) but degrades sharply on more complex ones (CIFAR-10) at high poisoning ratios, and its retention parameter can also over-reject benign clients.
{AlignIns}~\cite{xu2025detecting} exploits directional and sign-alignment signals, but like other detection-based methods remains sensitive to benign update variability under heterogeneity.
{SPMC}~\cite{he2025spmc} forms coalitions of mutually consistent clients and reweights updates according to coalition membership, but its consistency criterion degrades when benign clients are distributionally diverse, admitting stealthy adversaries that align with any sufficiently large coalition.

Overall, existing defenses either rely on restrictive assumptions on client distributions and adversarial behaviour, or introduce high FPRs that degrade main accuracy under realistic heterogeneous conditions.
These limitations motivate \textit{FedSurrogate}, a defense that accurately detects and mitigates backdoors while preserving benign knowledge across diverse data distributions.


\section{Threat Model and Design Goals}
\label{sec:threat_model_design_goals}

We consider a standard FL setting with a central server and a set of participating clients $\mathcal{N} = \{1,\dots,N\}$. In each communication round, the server distributes the current global model to the clients, each client performs local training on its private data, and returns a model update which the server aggregates. 

\textbf{Honest-majority assumption.} In line with previous studies~\cite{nguyen2022flame,qin2024resisting}, we assume an honest majority among clients: a subset $\mathcal{F}\subset\mathcal{N}$ may be adversarial but $|\mathcal{F}|<N/2$, so that benign clients collectively outweigh the malicious minority. We assume the server is honest-but-curious and uncompromised, and that adversaries cannot access or modify benign clients' local data or updates. We characterise the regime in which this assumption holds empirically in the Appendix {(see Fig.~\ref{fig:mcr_sensitivity_cifar10})}.

\textbf{Adversary model.} Malicious clients have full control over their local data and training procedure, can craft arbitrary model updates including scaled or maliciously modified parameters, and may coordinate with other malicious clients to mount distributed attacks.  
The adversary's objective is to induce the global model to misclassify trigger-bearing inputs to a selected target label while preserving main-task accuracy on clean data, so that the attack remains undetected by basic validation. We focus on targeted attacks that implant backdoors into the global model, and our evaluation considers three 
attack families that capture the primary mechanisms of trigger insertion, distributed coordination, and stealth-optimised evasion: \emph{CBA}~\cite{bagdasaryan2020backdoor}, \emph{DBA}~\cite{xie2019dba}; and \emph{Neurotoxin}~\cite{zhang2022neurotoxin} (see Sect.~\ref{sec:background}).


\noindent \textbf{Design goals of FedSurrogate.}
Given the adversary model, \textit{FedSurrogate} is designed to identify and neutralise malicious clients while maintaining global model utility. The defense pursues three primary goals: \emph{accurate identification} of malicious clients across all three attack families even under non-IID heterogeneity and low poisoning ratios; \emph{low false-positive rate} since incorrectly flagging an honest client whose data contribution would otherwise enrich the global model can meaningfully degrade utility; and \emph{preservation of main-task accuracy} both under attack and in non-adversarial rounds, ensuring that the defense behaves as a benign aggregation mechanism when no attack is present and does not introduce unnecessary disruption to the training process.

\section{Design of FedSurrogate}
\label{sec:method}
\begin{figure*}[t]
    \centering
    \makebox[\textwidth]{%
  \includegraphics[width=0.99\textwidth]{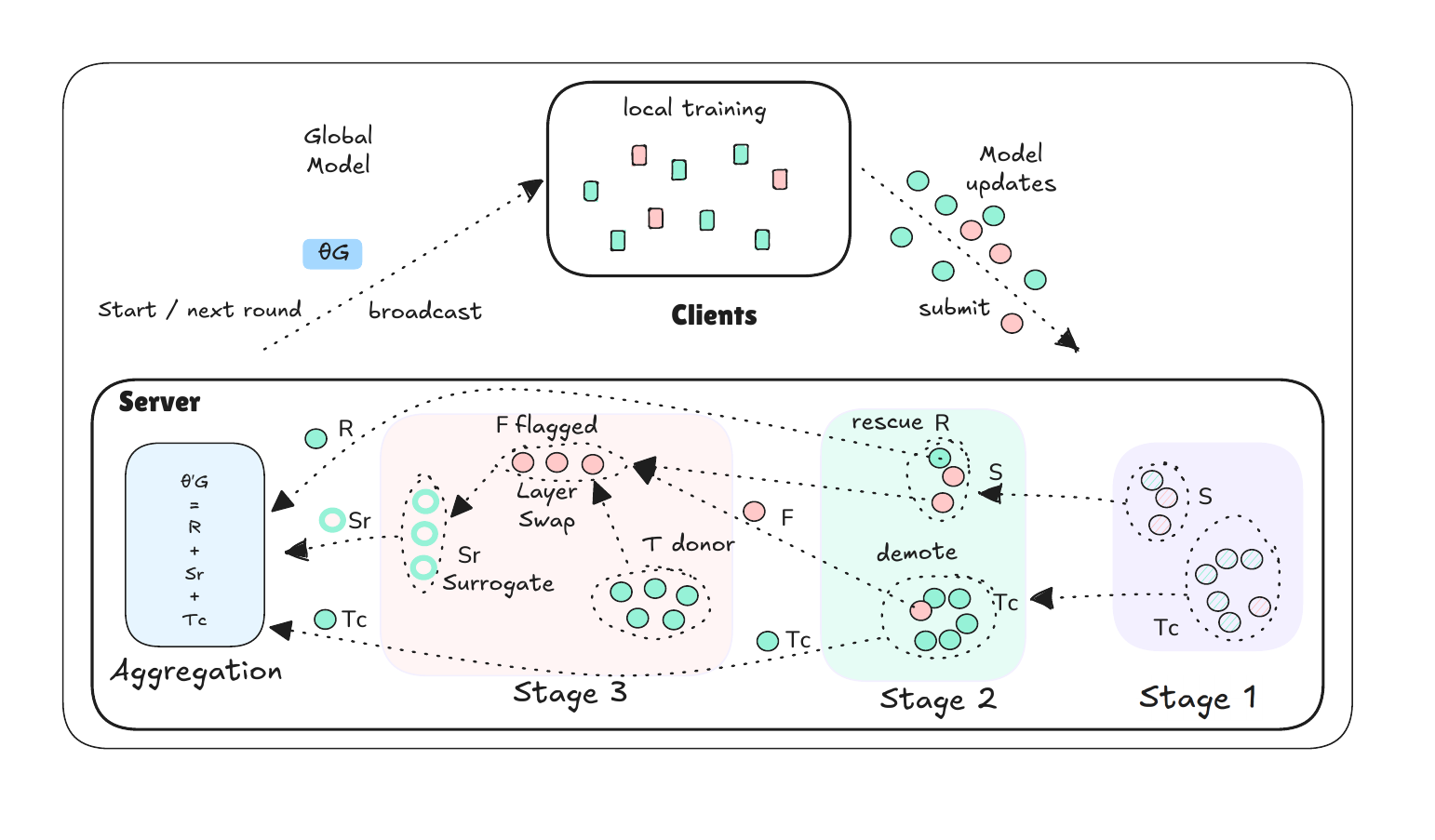}%
}
    \caption {Overview of the FedSurrogate defense pipeline. Stage 1 performs density-based clustering on layers identified via Layer Criticality Analysis, producing a coarse trusted set $\mathcal{T}_c$ and a suspect set $\mathcal{S}$. Stage 2 applies bidirectional gradient-alignment scoring to demote residual adversaries from $\mathcal{T}_c$ and rescue false positives $\mathcal{R}$ from $\mathcal{S}$. Stage 3 substitutes only the critical layers of confirmed malicious clients $\mathcal{F}$ with those of their nearest donor from $\mathcal{T}_c \cup \mathcal{R}$.}
    \label{fig:fedsurrogate}
\end{figure*}

We present \textit{FedSurrogate}, a multi-stage server-side defense that mitigates backdoor attacks in FL while explicitly addressing the over-rejection of benign clients under non-IID data heterogeneity which drives the high FPRs of existing detection-based defenses. The framework integrates three complementary stages: \emph{(i)} Layer Criticality Analysis and coarse clustering, \emph{(ii)} bidirectional gradient alignment filtering, and \emph{(iii)} surrogate replacement with controlled down-weighting. Figure~\ref{fig:fedsurrogate} illustrates the framework, and Algorithm~\ref{alg:fedsurrogate} presents the full procedure.


\vspace{0.2em}

\noindent \textbf{Stage 1: Layer Criticality Analysis and Coarse Clustering.}
\label{subsec:stage1-hdbscan}
After each communication round, the server receives updated models $\{\theta_i\}_{i=1}^N$ and computes per-client updates $\Delta_i = \theta_i - \theta_G$. Rather than clustering on entire model parameters, \textit{FedSurrogate} restricts clustering to a small set of \emph{security-critical layers} identified automatically at each round through Layer Criticality Analysis (LCA).

\textbf{Layer Criticality Analysis.} We observe that backdoor poisoning causes directional disagreement in specific layers: malicious clients push these layers towards the trigger-target mapping while benign clients do not. For each candidate layer $\ell$, we compute the mean pairwise cosine distance across client updates,
\begin{equation}
  d_\ell \;=\; \frac{2}{N(N-1)} \sum_{i<j} \left( 1 - \frac{\langle \Delta_i^\ell, \Delta_j^\ell \rangle}{\lVert \Delta_i^\ell \rVert_2 \, \lVert \Delta_j^\ell \rVert_2} \right),
\end{equation}
and normalise it relative to the per-round median, $\tilde d_\ell = d_\ell / \mathrm{median}(\{d_{\ell'}\})$. Since $\mathrm{median}(\tilde d) = 1$ by construction, layers form the \emph{critical layer set} when their normalised score exceeds
\begin{equation}
  L^* \;=\; \{\ell : \tilde d_\ell > 1 + \sigma \cdot \mathrm{MAD}(\tilde d)\},
\end{equation}
where $\sigma$ controls selection sensitivity. In practice, we select the top-$k$ layers by normalised divergence score ($k = 5$ in all experiments), which provides a stable selection size across rounds. Unlike prior work~\cite{nguyen2023fedgrad} that pre-selects a fixed layer, $L^*$ is recomputed dynamically per round and architecture, adapting to where adversarial signals concentrate under current attack and model state.


\textbf{Coarse clustering.} For each client, we form a feature vector $\mathbf{v}_i = \Delta_i[L^*]$ from updates restricted to the critical layers identified by LCA, and compute the pairwise cosine distance matrix $D_{ij} = 1 - \mathbf{v}_i^{\!\top} \mathbf{v}_j / (\lVert \mathbf{v}_i \rVert_2 \, \lVert \mathbf{v}_j \rVert_2)$. Cosine distance is chosen over Euclidean because it is invariant to per-client gradient magnitudes, which vary substantially under non-IID data volumes. We apply \textsc{HDBSCAN}~\cite{mcinnes2017hdbscan} to $D$, and under the honest-majority assumption the largest cluster $\mathcal{C}_{\max}$ (with $\lvert \mathcal{C}_{\max} \rvert \ge 0.5N + 1$) is taken as the \emph{coarse trusted set} $\mathcal{T}_c$, with the remaining clients forming the \emph{suspect set} $\mathcal{S}$. 
We label $\mathcal{T}_c$ as \emph{coarse} because under realistic non-IID conditions it may contain residual adversaries while $\mathcal{S}$ frequently contains misclassified benign clients; correcting both error types is the role of Stage~2.



\begin{algorithm}[t]
\caption{FedSurrogate} 
\label{alg:fedsurrogate}
\begin{algorithmic}[1]
\State \textbf{Input:}
\State \quad $\{\theta_i\}_{i=1}^N$, $\theta_G$, $\mathcal{H}$ \Comment{client models, global model, persistent client score memory}
\State \quad $L_r$, $\zeta$, $k$, $\gamma_s$, $\gamma_r$ \Comment{rescue layers, thresholds, shrink factors}
\State \textbf{Output:}
\State \quad $\theta_G'$,  $\mathcal{T}$,  $\mathcal{F}$ \Comment{Updated global model, trusted set, malicious set}

\State {\textbf{Stage 1: Layer Criticality Analysis and Coarse Clustering}}
\State \quad $\Delta \gets$ \Call{ClientUpdates}{$\{\theta_i\}, \theta_G$} \Comment{per-client deltas}
\State \quad $L^* \gets$ \Call{LCA}{$\Delta, k$} \Comment{layers with anomalous cross-client divergence}
\State \quad $\mathcal{T}_c \gets$ \Call{HDBSCAN}{$\Delta[L^*]$} \Comment{largest cluster = coarse trusted}
\State \quad $\mathcal{S} \gets [N] \setminus \mathcal{T}_c$

\State {\textbf{Stage 2: Bidirectional Gradient Alignment Filter}}
\State \quad $\hat{s} \gets$ \Call{AlignScore}{$\{\theta_i\}, \theta_G, L_r$} \Comment{cosine alignment on mid-deep layers}
\State \quad $\mathcal{H} \gets$ \Call{UpdateMemory}{$\mathcal{H}, \hat{s}, r$} \Comment{cumulative running average}

\State \quad \textbf{(2a) Screen trusted set: demote IQR outliers}
\State \quad $\mathcal{D} \gets$ \Call{IqrOutliers}{$\mathcal{H}[\mathcal{T}_c]$}
\State \quad $\mathcal{T}_c \gets \mathcal{T}_c \setminus \mathcal{D}$; \ $\mathcal{S} \gets \mathcal{S} \cup \mathcal{D}$

\State \quad \textbf{(2b) Rescue false positives from suspects}
\State \quad $\varepsilon \gets \min(\zeta,\ \text{median}(\mathcal{H}[\mathcal{S}]))$
\State \quad $\mathcal{R} \gets \{j \in \mathcal{S} : \mathcal{H}[j] \leq \varepsilon\}$
\State \quad $\mathcal{F} \gets \mathcal{S} \setminus \mathcal{R}$; \ $\mathcal{T} \gets \mathcal{T}_c \cup \mathcal{R}$

\State {\textbf{Stage 3: Surrogate Replacement}}
\State \quad $\mathcal{S}_r \gets \emptyset$ \Comment{surrogate-replaced set}
\ForAll{$f \in \mathcal{F}$}
    \State $d^* \gets$ \Call{NearestDonor}{$f, \mathcal{T}, L^*$}
    \State $\tilde\theta_f \gets$ \Call{ReplaceCriticalLayers}{$\theta_f, \theta_{d^*}, L^*$} \Comment{swap only $L^*$ from donor}
    \State $\mathcal{S}_r \gets \mathcal{S}_r \cup \{\tilde\theta_f\}$
\EndFor

\State {\textbf{Weighted Aggregation}}
\State \quad $\lambda_i \gets 1$ for $i \in \mathcal{T}_c$; $\ \gamma_s$ for $i \in \mathcal{R}$; $\ \gamma_r$ for $i \in \mathcal{S}_r$
\State \quad $\theta_G' \gets \dfrac{1}{\sum_i \lambda_i} \sum_{i \in \mathcal{T}_c \cup \mathcal{R} \cup \mathcal{S}_r} \lambda_i \, \tilde\theta_i$
\State \Return $\theta_G', \mathcal{T}, \mathcal{F}$
\end{algorithmic}
\end{algorithm}

\vspace{0.2em}

\noindent \textbf{Stage 2: Bidirectional Gradient Alignment Filter.}
\label{subsec:stage2-bidirectional-filter}
While LCA-driven clustering provides an effective first pass, it incurs substantial FPRs when client data distributions are heterogeneous, and at the same time sophisticated adversaries craft updates whose global statistics blend into the coarse trusted set. 
We thus introduce a \emph{bidirectional} filter that simultaneously screens $\mathcal{T}_c$ for residual contamination and rescues clients from $\mathcal{S}$ whose flagging is driven by heterogeneity rather than maliciousness.

\textbf{Mid-deep layer scoring.} Rather than reusing the dynamic set $L^*$ from Stage~1, gradient alignment scoring operates on a fixed, architecture-dependent set of \emph{mid-deep layers} $L_r$ (in ResNet-18, $L_r = \{\texttt{layer2.1}, \texttt{layer3}, \texttt{fc}\}$; in the simple CNN, $L_r = \{\texttt{fc1}, \texttt{fc2}\}$, the two network architectures we used). These  are chosen because they capture both trigger-level feature encodings and classification-level backdoor signals while remaining stable across rounds. This stability is essential: a fixed reference set ensures that cumulative alignment scores remain comparable across rounds, whereas the dynamic $L^*$ would introduce round-to-round noise into the temporal aggregation.


\textbf{Reference statistics and alignment score.} For each client $c_j$, we extract the mid-deep weights $\mathbf{w}_j = \theta_j[L_r]$ and compute the corresponding gradient $\mathbf{g}_j = \mathbf{w}_j - \theta_G[L_r]$. Reference statistics are computed over \emph{all clients}, not just suspects, using sample-size-weighted averaging $\mathbf{w}^* = \sum_i \omega_i \mathbf{w}_i$ and $\mathbf{g}^* = \sum_i \omega_i \mathbf{g}_i$, where $\omega_i = n_i / \sum_j n_j$ and $n_i$ is the sample count at client $i$. Computing references over the full population prevents Byzantine coordination from biasing the baseline. We then define the instantaneous alignment score
\begin{equation}
  \hat{s}_j \;=\; \frac{(\mathbf{w}_j - \mathbf{w}^*)^{\!\top} \mathbf{g}^*}{\lVert \mathbf{w}_j - \mathbf{w}^* \rVert_2 \cdot \lVert \mathbf{g}^* \rVert_2},
  \label{eq:alignment-score}
\end{equation}
which is high for benign clients whose loss-minimisation objectives broadly align with $\mathbf{g}^*$ and low for adversaries pursuing conflicting objectives. Scores are normalised to $[0,1]$ via min-max scaling and aggregated across rounds
\begin{equation}
  s_j^{(r)} \;=\; \frac{m_j - 1}{m_j} \cdot s_j^{(r-1)} \;+\; \frac{1}{m_j} \cdot \hat{s}_j^{(r)},
  \label{eq:ema-score}
\end{equation}
where $m_j$ counts the rounds $c_j$ has been scored. Temporal aggregation distinguishes persistent adversarial behaviour from transient heterogeneity-induced anomalies, allowing benign clients misclustered in a single round to accumulate evidence of their legitimacy.

\textbf{Bidirectional correction.} To detect residual adversaries that evaded Stage~1, we apply an inter-quartile range (IQR) test to the cumulative scores of coarse trusted clients: clients whose score exceeds the upper fence $\tau_{\mathrm{scr}} = q_3 + 1.5(q_3 - q_1)$ are demoted from $\mathcal{T}_c$ to $\mathcal{S}$. To rescue false positives among suspects, we compute an adaptive cutoff
\begin{equation}
  \varepsilon \;=\; \min \big( \zeta,\, \mathrm{median}(\{s_j : j \in \mathcal{S}\}) \big),
\end{equation}
where $\zeta$ caps the threshold to prevent it from drifting upward when an adversary attempts to pollute the suspect pool with benign-looking updates. Clients satisfying $s_j \le \varepsilon$ are rescued, yielding $\mathcal{R} = \{j \in \mathcal{S} : s_j \le \varepsilon\}$, $\mathcal{F} = \mathcal{S} \setminus \mathcal{R}$, and $\mathcal{T} = \mathcal{T}_c \cup \mathcal{R}$. 
Thus, Stage~2 partitions $\mathcal{S}$ into a rescued set $\mathcal{R}$ (false positives returned to the trusted pool) and a confirmed-malicious set $\mathcal{F}$. The median-based threshold is adaptive: under a strong attack the suspect pool contains mostly true adversaries and the median is high, making rescue conservative; under benign heterogeneity the median is low and rescue is permissive. 


\vspace{0.2em}

\noindent \textbf{Stage 3: Surrogate Replacement.}
\label{subsec:stage3-replacement}
The confirmed-malicious set $\mathcal{F}$ from Stage~2 is passed to Stage~3, where each flagged client is replaced with a \emph{surrogate} drawn from the trusted set $\mathcal{T} = \mathcal{T}_c \cup \mathcal{R}$, producing a surrogate-replaced set $\mathcal{S}_r = \{\tilde{\theta}_f : f \in \mathcal{F}\}$ that participates in aggregation in place of $\mathcal{F}$. This keeps all $N$ clients contributing to aggregation, avoiding the reduction in effective sample size that outright exclusion would impose.

\textbf{Donor selection and surrogate replacement.} For each $f \in \mathcal{F}$, a donor is selected from $\mathcal{T}$ by nearest-neighbour search on the Stage~1 distance matrix, $d^*(f) = \argmin_{d \in \mathcal{T}} D[f, d]$, ensuring that the donor has parameter structure similar to the flagged client. Rather than replacing the entire model, which discards legitimate local knowledge, we perform a \emph{layer-selective replacement} that substitutes only the critical layers $L^*$:
\begin{equation}
  \tilde{\theta}_f^\ell \;=\;
  \begin{cases}
    \theta_{d^*(f)}^\ell & \text{if } \ell \in L^*, \\[4pt]
    \theta_f^\ell & \text{otherwise}.
  \end{cases}
  \label{eq:surrogate-construction}
\end{equation}
As $L^*$ is the same set on which clustering and donor selection operate, the replacement targets precisely the layers where the adversarial signal concentrates, neutralising the backdoor while preserving the flagged client's non-critical contributions.


\noindent\textbf{Differential aggregation weights.} To mitigate the risk that adversaries craft updates close to benign gradients~\cite{fang2020local,bagdasaryan2020backdoor}, we apply differential weights to the 
sets:
\begin{equation}
  \lambda_i \;=\; 
  \begin{cases}
    1.0 & \text{if } i \in \mathcal{T}_c, \\
    \gamma_s & \text{if } i \in \mathcal{R}, \\
    \gamma_r & \text{if } i \in \mathcal{S}_r,
  \end{cases}
  \label{eq:shrink-weights}
\end{equation}
with $\gamma_s = 0.7$ for rescued clients and $\gamma_r = 0.3$ for surrogate-replaced clients, reflecting residual uncertainty from Stage~1 flagging and bounding contributions when donor selection is suboptimal, respectively. The global model is updated:
\begin{equation}
\theta_G' \;=\;
\frac{1}{\sum_{i} \lambda_i}
\sum_{i \in \mathcal{T}_c \cup \mathcal{R} \cup \mathcal{S}_r} \lambda_i \, \tilde{\theta}_i,
\label{eq:weighted-aggregation}
\end{equation}
where $\tilde{\theta}_i = \theta_i$ for $i \in \mathcal{T}_c \cup \mathcal{R}$ and $\tilde{\theta}_i$ is constructed via Eq.~\eqref{eq:surrogate-construction} for $i \in \mathcal{S}_r$.


\section{Experimental Setup and Results}



\textbf{Datasets and model architectures.}
\label{subsec:datasets}
We evaluate on four image-classification benchmarks of increasing complexity: {MNIST}~\cite{deng2012mnist}, {Fashion-MNIST}~\cite{xiao2017fashion}, {CIFAR-10}~\cite{krizhevsky2009learning}, and {CIFAR-100}~\cite{krizhevsky2009learning}. To simulate heterogeneous client data distributions typical of FL, we partition each dataset using a Dirichlet distribution with concentration parameter $\alpha=0.5$, in line with prior work~\cite{nguyen2022flame,nguyen2023fedgrad,qin2024resisting}. For MNIST and Fashion-MNIST we use a 4-layer CNN (2 convolutional layers and 2 fully-connected layers); for CIFAR-10 and CIFAR-100 we adopt ResNet-18~\cite{he2016deep} to handle the increased complexity of natural images.

\textbf{FL configuration and hyperparameters.}
\label{subsec:hyperparams}
All experiments adopt synchronous FL with one central server and $N = 20$ clients, each participating in every communication round. Benign clients perform $E = 2$ local epochs of SGD, while malicious clients perform $E = 5$ epochs in line with prior backdoor evaluation practice~\cite{bagdasaryan2020backdoor,xie2019dba}. Learning rates are $\eta = 0.01$ for the CNN and $\eta = 0.1$ for ResNet-18, with batch size $b = 32$, and $R = 100$ communication rounds. \footnote{Our code is public and available at: \url{https://github.com/fabacha/FedSurrogate-Backdoor-Defense-in-FL.git}} 

\textbf{Backdoor attacks and adversary configuration.}
\label{subsec:attacks-adversary}
We evaluate under three backdoor attack families: \emph{(i)} CBA~\cite{bagdasaryan2020backdoor}, where each malicious client injects the full trigger; \emph{(ii)} DBA~\cite{xie2019dba}, where the trigger is decomposed across multiple adversaries; and \emph{(iii)} Neurotoxin~\cite{zhang2022neurotoxin}, a stealth-optimised attack. 
Triggers follow established patterns: for CIFAR-10/100, a $3 \times 3$ red square in the lower-right corner with target label \emph{automobile} (class 1); for MNIST/Fashion-MNIST, a $3 \times 3$ white patch with target label 1. Unless otherwise specified, MCR\,$=0.2$ and each malicious client poisons $30\%$ of its local data ($pdr = 0.3$). 

\textbf{Baseline defenses and evaluation metrics.}
\label{subsec:baselines-metrics}
We compare \textit{FedSurrogate} against seven state-of-the-art defenses: {FoolsGold}~\cite{fung2018mitigating}, {FLAME}~\cite{nguyen2022flame}, {FedGrad}~\cite{nguyen2023fedgrad}, {Snowball}$\boxminus$\footnote{We use {Snowball}$\boxminus$ to denote the variant of Snowball without the VAE warm-up stage, adopted to fit our $R=100$ round budget.}~\cite{qin2024resisting}, {AlignIns}~\cite{xu2025detecting}, {FLShield}~\cite{kabir2024flshield}, and {SPMC}~\cite{he2025spmc}. 
 We report four primary metrics: \emph{Main-task Accuracy (MTA)}, the top-1 accuracy on the clean test set; \emph{Attack Success Rate (ASR)}, the proportion of trigger-bearing test inputs misclassified to the target label (lower is better); \emph{True-Positive Rate (TPR)}, the proportion of malicious clients correctly detected; and \emph{False-Positive Rate (FPR)}, the proportion of benign clients incorrectly flagged (lower is better).

\textbf{Baseline validation on MNIST and Fashion-MNIST.}
Table~\ref{tab:main_results_mnist} reports MTA and ASR for all defenses on MNIST and Fashion-MNIST, two benchmarks of increasing complexity that serve as initial validation rather than a strong test of robustness, consistent with prior observations that simple datasets tend to overestimate defense performance~\cite{khan2023pitfalls}. On MNIST, most defenses mitigate all three attacks, with average ASR well below one percent and MTA preserved within a fraction of a point of FedAvg. 
On Fashion-MNIST, where non-IID variation under Dirichlet partitioning is more pronounced, meaningful differences emerge: FoolsGold and AlignIns show sharply elevated ASR, indicating difficulty in distinguishing benign heterogeneity from adversarial deviation; SPMC fails to defend across both datasets; FLShield mitigates the centralised attack but collapses on the stealthier distributed and Neurotoxin variants; and FLAME, FedGrad, and Snowball$\boxminus$ achieve low ASR but at the cost of 3--8 percentage point MTA drops due to excessive benign filtering. 
\textit{FedSurrogate} achieves the highest average MTA among effective defenses on both datasets (99.10\% on MNIST, 88.78\% on Fashion-MNIST) while maintaining average ASR at 0.09\% and 0.16\% respectively, introducing no measurable utility loss on MNIST and remaining within 1.27 points of FedAvg on Fashion-MNIST. These results confirm that the robustness of \textit{FedSurrogate} is structural rather than dataset-specific.

\begin{table*}[t]
\centering
\caption{MTA and ASR results of baselines and \textit{FedSurrogate} on non-IID MNIST and Fashion-MNIST.
Results are shown in \%. All experiments use $pdr{=}0.3$ on non-IID partitions ($\alpha{=}0.5$).
The best result in each column is \textbf{bold} and the second best is \underline{underlined}, considering only defenses that achieve both: MTA within $5\%$ of FedAvg's MTA, and successfully mitigate the attacks (ASR $< 5\%$), marked in \colorbox{lightgray}{light grey}.
$\Delta$ row shows absolute change relative to FedAvg.}
\label{tab:main_results_mnist}
\resizebox{\textwidth}{!}{%
\renewcommand{\arraystretch}{1.15}
\begin{tabular}{@{}ll|cc|cc|cc|cc@{}}
\toprule
\scriptsize
\multirow{2}{*}{\textbf{Dataset}}
  & \multirow{2}{*}{\textbf{Methods}}
  & \multicolumn{2}{c|}{\textbf{CBA}}
  & \multicolumn{2}{c|}{\textbf{DBA}}
  & \multicolumn{2}{c|}{\textbf{Neurotoxin}}
  & \multirow{2}{*}{\textbf{Avg. MTA$\uparrow$}}
  & \multirow{2}{*}{\textbf{Avg. ASR$\downarrow$}} \\
\cmidrule(lr){3-4}\cmidrule(lr){5-6}\cmidrule(lr){7-8}
  &
  & MTA$\uparrow$ & ASR$\downarrow$
  & MTA$\uparrow$ & ASR$\downarrow$
  & MTA$\uparrow$ & ASR$\downarrow$
  & & \\
\midrule
\multirow{11}{*}{\rotatebox[origin=c]{90}{\textbf{MNIST}}}
  & FedAvg        & 99.15 & 99.97 & 99.17 & 99.20 & 99.15 & 99.90 & 99.16 & 99.69 \\
\cmidrule(lr){2-10}
&  FoolsGold~\cite{fung2018mitigating}     & \cellcolor{lightgray} 98.79 & \cellcolor{lightgray} 0.16  &  98.84 &  7.12  & \cellcolor{lightgray} 98.80 & \cellcolor{lightgray} 0.62  & \cellcolor{lightgray} 98.81 & \cellcolor{lightgray} 2.63 \\
  & FedGrad~\cite{nguyen2023fedgrad}       & \cellcolor{lightgray} 97.12 & \cellcolor{lightgray} 0.34  & \cellcolor{lightgray} 97.82 & \cellcolor{lightgray} 0.38  & 91.98 & 99.84 &  95.64 & 33.52 \\
  & FLAME~\cite{nguyen2022flame}         & \cellcolor{lightgray} \underline{98.91} & \cellcolor{lightgray} 0.12  & \cellcolor{lightgray} \underline{98.90} & \cellcolor{lightgray} 0.12  & \cellcolor{lightgray} 98.88 & \cellcolor{lightgray} 0.14  & \cellcolor{lightgray} \underline{98.90} & \cellcolor{lightgray} 0.13 \\
  & FLShield~\cite{kabir2024flshield}      & 99.14 & 99.92 & 99.23 & 99.04 & 99.14 & 99.92 & 99.17 & 99.63 \\
  & Snowball$\boxminus$~\cite{qin2024resisting}      & \cellcolor{lightgray} 98.83 & \cellcolor{lightgray} \textbf{0.05}  & \cellcolor{lightgray} 98.89 & \cellcolor{lightgray} \textbf{0.06}  & \cellcolor{lightgray} 98.78 & \cellcolor{lightgray} \underline{0.11}  & \cellcolor{lightgray} {98.83} & \cellcolor{lightgray} \textbf{0.07} \\
  & AlignIns~\cite{xu2025detecting}      & 99.11 & 99.93 & {99.03} & 5.12  & \cellcolor{lightgray} \underline{99.02} & \cellcolor{lightgray} \textbf{0.07}  &  {99.05} &  35.04 \\
  & SPMC~\cite{he2025spmc}          & 99.28 & 99.97 & 99.25 & 99.24 & 99.22 & 99.93 & 99.25 & 99.71 \\
\cmidrule(lr){2-10}
  & \textbf{FedSurrogate}
  & \cellcolor{lightgray} \textbf{99.11} & \cellcolor{lightgray} \underline{0.09}  & \cellcolor{lightgray} \textbf{99.13} & \cellcolor{lightgray} \underline{0.08}
  & \cellcolor{lightgray} \textbf{99.07} & \cellcolor{lightgray} \underline{0.11}  & \cellcolor{lightgray} \textbf{99.10} & \cellcolor{lightgray} \underline{0.09} \\
  & {\scriptsize $\Delta$ vs FedAvg}
  & {\scriptsize $\downarrow$0.04} & {\scriptsize $\downarrow$99.88}
  & {\scriptsize $\downarrow$0.04} & {\scriptsize $\downarrow$99.12}
  & {\scriptsize $\downarrow$0.08} & {\scriptsize $\downarrow$99.79}
  & {\scriptsize $\downarrow$0.06} & {\scriptsize $\downarrow$99.60} \\
\midrule
\multirow{11}{*}{\rotatebox[origin=c]{90}{\textbf{Fashion-MNIST}}}
  & FedAvg        & 90.37 & 97.71 & 90.08 & 96.57 & 89.70 & 97.03 & 90.05 & 97.10 \\
\cmidrule(lr){2-10}
  
& FoolsGold~\cite{fung2018mitigating}     & 90.28 & 97.81 & 90.08 & 97.19 & 89.75 & 97.41 & 90.04 & 97.47 \\
  & FedGrad~\cite{nguyen2023fedgrad}       &  82.45 &  0.26  &  82.41 &  0.26  &  82.44 &  0.26  &  82.43 &  0.26 \\
  & FLAME~\cite{nguyen2022flame}         & \cellcolor{lightgray} 86.91 & \cellcolor{lightgray} \underline{0.22}  & \cellcolor{lightgray} \underline{86.94} & \cellcolor{lightgray} \underline{0.22}  & \cellcolor{lightgray} \underline{86.52} & \cellcolor{lightgray} \underline{0.23}  & \cellcolor{lightgray} \underline{86.79} & \cellcolor{lightgray} \underline{0.22} \\
  & FLShield ~\cite{kabir2024flshield}     & \cellcolor{lightgray} \textbf{89.21} & \cellcolor{lightgray} 0.30  & 89.45 & 93.42 & 91.32    & 98.80    & 89.99    & 64.17 \\
  & Snowball$\boxminus$~\cite{qin2024resisting}      & 81.18 &  0.28  &  70.19 &  4.69  &  73.71 &  {0.14}  &  75.03 &  1.70 \\
  & AlignIns~\cite{xu2025detecting}      & 86.75 & 95.42 & 87.32 & 95.00 & 86.67 & 95.93 & 86.91 & 95.45 \\
  & SPMC~\cite{he2025spmc}          & 90.03 & 98.16 & 89.74 & 97.16 & 91.31 & 98.63 & 90.36 & 97.98 \\
\cmidrule(lr){2-10}
  & \textbf{FedSurrogate}
  & \cellcolor{lightgray} \underline{88.99} & \cellcolor{lightgray} \textbf{0.17}  & \cellcolor{lightgray} \textbf{88.99} & \cellcolor{lightgray} \textbf{0.18}
  & \cellcolor{lightgray} \textbf{88.35} & \cellcolor{lightgray} \textbf{0.14}  & \cellcolor{lightgray} \textbf{88.78} & \cellcolor{lightgray} \textbf{0.16} \\
  & {\scriptsize $\Delta$ vs FedAvg}
  & {\scriptsize $\downarrow$1.38} & {\scriptsize $\downarrow$97.54}
  & {\scriptsize $\downarrow$1.09} & {\scriptsize $\downarrow$96.39}
  & {\scriptsize $\downarrow$1.35} & {\scriptsize $\downarrow$96.89}
  & {\scriptsize $\downarrow$1.27} & {\scriptsize $\downarrow$96.94} \\
\bottomrule
\end{tabular}%
}
\end{table*}



\begin{table*}[t]
\centering
\caption{MTA and ASR results of baselines and \textit{FedSurrogate} on non-IID CIFAR-10 and CIFAR-100.
Results are shown in \%. All experiments use $pdr{=}0.3$ with ResNet-18 on non-IID partitions ($\alpha{=}0.5$).
The best result in each column is \textbf{bold} and the second best is \underline{underlined}, considering only defenses that achieve both: MTA within $5\%$ of FedAvg's MTA, and successfully mitigate the attacks (ASR $< 5\%$), marked in \colorbox{lightgray}{light grey}.
$\Delta$ row shows absolute change relative to FedAvg.}
\label{tab:main_results_cifar}
\resizebox{\textwidth}{!}{%
\renewcommand{\arraystretch}{1.15}
\begin{tabular}{@{}ll|cc|cc|cc|cc@{}}
\toprule
\scriptsize
\multirow{2}{*}{\textbf{Dataset}}
  & \multirow{2}{*}{\textbf{Methods}}
  & \multicolumn{2}{c|}{\textbf{CBA}}
  & \multicolumn{2}{c|}{\textbf{DBA}}
  & \multicolumn{2}{c|}{\textbf{Neurotoxin}}
  & \multirow{2}{*}{\textbf{Avg. MTA$\uparrow$}}
  & \multirow{2}{*}{\textbf{Avg. ASR$\downarrow$}} \\
\cmidrule(lr){3-4}\cmidrule(lr){5-6}\cmidrule(lr){7-8}
  &
  & MTA$\uparrow$ & ASR$\downarrow$
  & MTA$\uparrow$ & ASR$\downarrow$
  & MTA$\uparrow$ & ASR$\downarrow$
  & & \\
\midrule
\multirow{10}{*}{\rotatebox[origin=c]{90}{\textbf{CIFAR-10}}}
  & FedAvg        & 90.25 & 99.77 & 90.15 & 46.42 & 89.74 & 99.53 & 90.05 & 81.91 \\
\cmidrule(lr){2-10}
  & FoolsGold~\cite{fung2018mitigating}     & 87.34 & 70.43 &  85.55 &  5.71  & \cellcolor{lightgray} \underline{86.13} & \cellcolor{lightgray} \textbf{1.48}  & {86.34} &  25.87 \\
  & FedGrad~\cite{nguyen2023fedgrad}       &  80.88 &  {0.86}  &  82.08 &  {0.88}  & 82.26 & 99.93 &  81.74 &  33.89 \\
  & FLAME~\cite{nguyen2022flame}         & {83.24} &  1.91  &  83.19 &  2.12  &  {83.48} &  {1.33}  &  {83.30} & {1.79} \\
  & FLShield~\cite{kabir2024flshield}      & 90.05 & 99.74 & {89.86} &  28.62 & 89.66 & 99.57 & 89.86 & 75.98 \\
  & Snowball$\boxminus$~\cite{qin2024resisting}      &  69.85 &  5.03  & 60.52 & {0.42}  & 34.21 & 9.78  &  54.86 & 5.08 \\
  & AlignIns~\cite{xu2025detecting}      & 87.79 & 87.76 & \cellcolor{lightgray} \underline{87.75} & \cellcolor{lightgray} \textbf{1.03}  & 87.23 & 93.23 & 87.59 & 60.67 \\
  & SPMC~\cite{he2025spmc}          & 90.18 & 99.51 & {89.67} &  38.93 & 89.30 & 99.76 & {89.72} & 79.40 \\
\cmidrule(lr){2-10}
  & \textbf{FedSurrogate}
  & \cellcolor{lightgray} \textbf{87.92}
  & \cellcolor{lightgray} \textbf{1.59}
  & \cellcolor{lightgray} \textbf{88.10}
  & \cellcolor{lightgray} \underline{1.66}
  & \cellcolor{lightgray} \textbf{86.93}
  & \cellcolor{lightgray} \underline{2.06}
  & \cellcolor{lightgray} \textbf{87.65}
  & \cellcolor{lightgray} \textbf{1.77} \\
  & {\scriptsize $\Delta$ vs FedAvg}
  & {\scriptsize $\downarrow$2.33}
  & {\scriptsize $\downarrow$98.18}
  & {\scriptsize $\downarrow$2.05}
  & {\scriptsize $\downarrow$44.76}
  & {\scriptsize $\downarrow$2.81}
  & {\scriptsize $\downarrow$97.47}
  & {\scriptsize $\downarrow$2.40}
  & {\scriptsize $\downarrow$80.14} \\
\midrule
\multirow{10}{*}{\rotatebox[origin=c]{90}{\textbf{CIFAR-100}}}
  & FedAvg        & 69.84 & 99.03 & 69.36 & 34.41 & 70.04 & 96.43 & 69.75 & 76.62 \\
\cmidrule(lr){2-10}
  & FoolsGold~\cite{fung2018mitigating}     &  64.52 &  0.27  & {67.44} &  10.35 & \cellcolor{lightgray} \textbf{67.28} & \cellcolor{lightgray} \textbf{0.19}  & \cellcolor{lightgray} \underline{66.41} & \cellcolor{lightgray} \underline{3.60} \\
  & FedGrad~\cite{nguyen2023fedgrad}       &  60.75 &  {0.13}  &  59.68 &  0.28  & 44.64 & 99.97 &  55.02 &  33.46 \\
  & FLAME~\cite{nguyen2022flame}         &  56.20 &  0.48  &  57.03 &  0.26  &  57.01 & {0.12}  &  56.75 &  0.29 \\
  & FLShield~\cite{kabir2024flshield}      & 69.30 & 95.75 &  {69.52} &  34.91 & 70.25 & 88.64 & 69.69 & 73.10 \\
  & Snowball$\boxminus$~\cite{qin2024resisting}      & 47.42 & 0.67  & 47.14 & 0.09  & 40.24 & 0.34  & 44.93 & 0.37 \\
  & AlignIns~\cite{xu2025detecting}      & \cellcolor{lightgray} \underline{66.45} & \cellcolor{lightgray} \textbf{0.18}  & \cellcolor{lightgray} \underline{66.39} & \cellcolor{lightgray} \textbf{0.15}  &  63.49 & 0.20  &  65.44 & {0.18} \\
  & SPMC~\cite{he2025spmc}          & 69.72 & 96.83 & {69.52} &  24.67 & 70.55 & 96.38 & {69.93} & 72.63 \\
\cmidrule(lr){2-10}
  & \textbf{FedSurrogate}
  & \cellcolor{lightgray} \textbf{66.66}
  & \cellcolor{lightgray} \underline{0.24}
  & \cellcolor{lightgray} \textbf{66.58}
  & \cellcolor{lightgray} \underline{0.18}
  & \cellcolor{lightgray} \underline{66.81}
  & \cellcolor{lightgray} \underline{0.31}
  & \cellcolor{lightgray} \textbf{66.68}
  & \cellcolor{lightgray} \textbf{0.24} \\
  & {\scriptsize $\Delta$ vs FedAvg}
  & {\scriptsize $\downarrow$3.18}
  & {\scriptsize $\downarrow$98.79}
  & {\scriptsize $\downarrow$2.78}
  & {\scriptsize $\downarrow$34.23}
  & {\scriptsize $\downarrow$3.23}
  & {\scriptsize $\downarrow$96.12}
  & {\scriptsize $\downarrow$3.07}
  & {\scriptsize $\downarrow$76.38} \\
\bottomrule
\end{tabular}%
}

\end{table*}


\textbf{CIFAR-10 and CIFAR-100: realistic image classification.}
Table~\ref{tab:main_results_cifar} reports MTA and ASR on CIFAR-10 and CIFAR-100, evaluated with ResNet-18 under the same non-IID partitioning. On CIFAR-10, \textit{FedSurrogate} achieves both the highest average MTA and the lowest average ASR among all defenses at 87.65\% and 1.77\% respectively. On CIFAR-100, where the expanded label space sharpens the non-IID effect, \textit{FedSurrogate} maintains the highest average MTA at 66.68\% with the second-lowest average ASR at 0.24\%, narrowly behind AlignIns at 0.18\%. Notably, AlignIns shows a marked improvement on CIFAR-100 compared to its selective failures on the simpler datasets, which we attribute to the wider diversity of benign update directions under fine-grained class partitioning making tightly clustered malicious updates easier to isolate; however, AlignIns pays a 1.24-point MTA penalty relative to \textit{FedSurrogate}.

Among the remaining baselines, FLAME is the only defense that consistently suppresses the backdoor across both datasets, but incurs substantial MTA reductions of 4.35 and 9.93 percentage points on CIFAR-10 and CIFAR-100 respectively due to over-filtering. FLShield and SPMC fail to defend on both datasets, with average ASR exceeding 72\%, while FedGrad collapses entirely on Neurotoxin with ASR reaching 99.93--99.97\%, and Snowball$\boxminus$ over-filters to the point of rendering the global model unusable, with MTA dropping to 54.86\% on CIFAR-10 and 44.93\% on CIFAR-100. Together, this confirms that \textit{FedSurrogate} is the only defense that consistently maintains both high MTA and low ASR across all four datasets, with the gap to competing methods widening as task complexity increases.

\textbf{Detection performance across all datasets.}
Table~\ref{tab:detection_digits} reports the per-defense ability to correctly classify malicious and benign clients across all four datasets and three attacks. \textit{FedSurrogate} achieves consistently high TPR (98.8-100\%) while maintaining the lowest FPR among all defenses on every dataset-attack combination, with FPR values of 1.1-1.2\% on the MNIST family and 2.3-9.1\% on the CIFAR family, substantially below those of every other effective baseline. The remaining defenses fall into two failure regimes that persist across datasets: methods such as FedGrad, FLAME, and Snowball$\boxminus$ achieve high TPR but at FPR values exceeding 31\%, 48\%, and 87\% respectively, indicating that they reject the majority of benign clients alongside the true adversaries; while FLShield, AlignIns, and SPMC fail to detect adversaries reliably, with TPR values frequently below 35\% on the MNIST family and only AlignIns recovering on CIFAR-100 for the reasons discussed earlier. FedGrad additionally collapses entirely on Neurotoxin across both CIFAR datasets, with TPR dropping to 1.0\% and FPR reaching 98--100\%, mirroring its ASR collapse in Table~\ref{tab:main_results_cifar} and reinforcing that detection signals tuned to conventional backdoor patterns generalise poorly to stealth-optimised attacks. It is worth noting that \textit{FedSurrogate} attains TPR values slightly below 100\% in several settings, most notably 98.8\% on Neurotoxin under CIFAR-100 and 99.0\% on Neurotoxin under CIFAR-10, yet still achieves the lowest or near-lowest ASR in those same settings, as reported in Tables~\ref{tab:main_results_mnist} and~\ref{tab:main_results_cifar}. We attribute this to the Layer Criticality Analysis stage, which restricts both clustering and gradient-alignment scoring to the small subset of layers carrying the backdoor signal, so that the rare malicious update slipping through detection in one round contributes a backdoor signal that is too weak to accumulate meaningfully across the training trajectory, particularly once subsequent rounds catch and replace the adversary. Together, these results confirm that \textit{FedSurrogate} maintains a high-utility global model while admitting the overwhelming majority of honest clients into aggregation, reducing the systematic exclusion of benign participants that characterises the more aggressive baseline defenses.

\begin{figure}[H]
  \centering
  \includegraphics[width=0.85\textwidth]{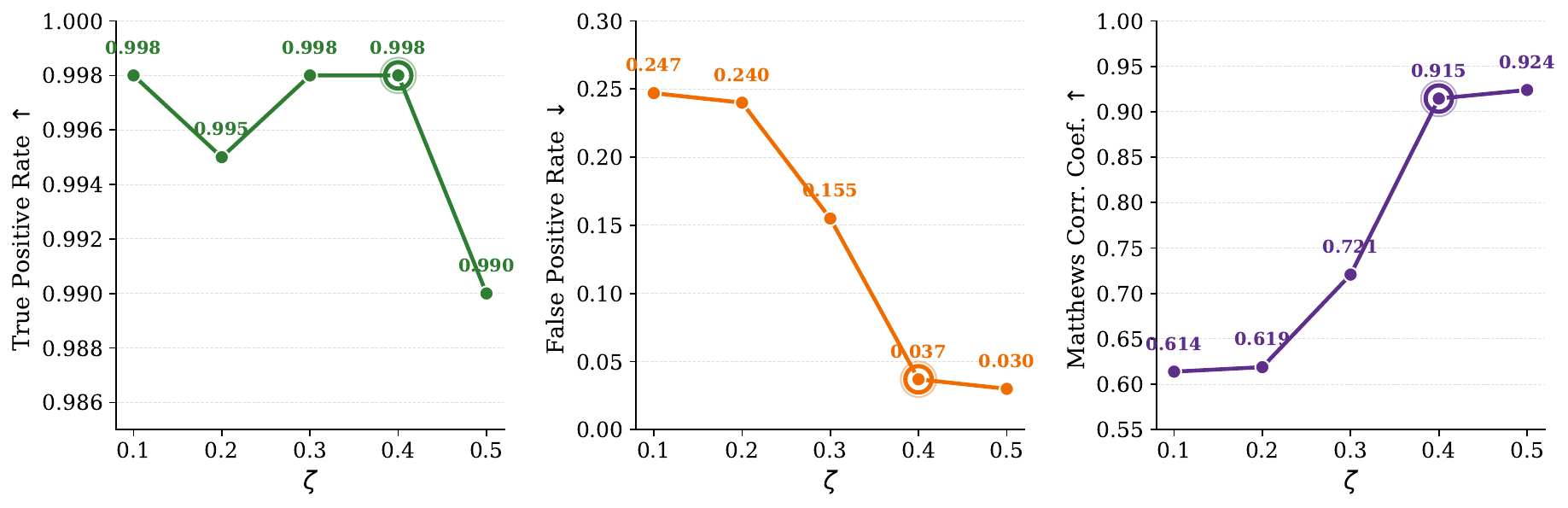}
  \caption{Sensitivity of \textit{FedSurrogate}'s detection performance to the threshold $\zeta$ on CIFAR-10 ($pdr{=}0.3$).} 
 
  \label{fig:zeta_ablation}
\end{figure}

\textbf{Sensitivity to the threshold parameter \texorpdfstring{$\zeta$}{zeta}.}
\label{sec:zeta_ablation}
The threshold $\zeta$ governs how aggressively \textit{FedSurrogate} discriminates between benign and malicious clients during the bidirectional gradient alignment stage. To assess sensitivity and justify our chosen value, we sweep $\zeta \in \{0.1, 0.2, 0.3, 0.4, 0.5\}$ on CIFAR-10 with $pdr{=}0.3$ and report TPR, FPR, and Matthews Correlation Coefficient (MCC), the latter providing a more reliable single summary than F1 for imbalanced binary classification~\cite{chicco2020advantages}. As shown in Fig.~\ref{fig:zeta_ablation}, TPR remains at its maximum of 0.998 for $\zeta \in \{0.1, 0.3, 0.4\}$ and declines only at $\zeta = 0.5$, while FPR falls steadily from 0.247 at $\zeta = 0.1$ to 0.037 at $\zeta = 0.4$. MCC exhibits a clear elbow at $\zeta = 0.4$, rising sharply from 0.721 at $\zeta = 0.3$ to 0.915 before plateauing at 0.924 for $\zeta = 0.5$. We therefore select $\zeta = 0.4$ as the operating point: it is the largest threshold that retains the maximum observed TPR, yields a near-optimal MCC within 0.009 of the maximum, and reduces FPR by 76\% relative to $\zeta = 0.3$.



\begin{table}[H]
\centering
\caption{Detection performance on non-IID settings. All experiments use $pdr{=}0.3$, non-IID ($\alpha{=}0.5$). The best result in each column is \textbf{bold} and the second best is \underline{underlined}, considering only defenses that have TPR $> 90\%$ and FPR $< 10\%$, marked in \colorbox{lightgray}{light grey}.}
\label{tab:detection_digits}
\setlength{\tabcolsep}{3.5pt}
\renewcommand{\arraystretch}{1.15}
\scriptsize
\begin{tabular}{@{}ll|cc|cc|cc@{}}
\toprule
\multirow{2}{*}{\textbf{Dataset}}
  & \multirow{2}{*}{\textbf{Method}}
  & \multicolumn{2}{c|}{\textbf{CBA}}
  & \multicolumn{2}{c|}{\textbf{DBA}}
  & \multicolumn{2}{c}{\textbf{Neurotoxin}} \\
\cmidrule(lr){3-4}\cmidrule(lr){5-6}\cmidrule(lr){7-8}
  &
  & TPR$\uparrow$ & FPR$\downarrow$
  & TPR$\uparrow$ & FPR$\downarrow$
  & TPR$\uparrow$ & FPR$\downarrow$ \\
\midrule

\multirow{9}{*}{\rotatebox[origin=c]{90}{\textbf{MNIST}}}
  & FoolsGold~\cite{fung2018mitigating}\textsuperscript{$\dagger$}
                  & \cellcolor{lightgray} \underline{99.5}   & \cellcolor{lightgray} \underline{7.7}   & \cellcolor{lightgray} \underline{97.5}   & \cellcolor{lightgray} \underline{7.5}   & \cellcolor{lightgray} \underline{98.0}   & \cellcolor{lightgray} \underline{7.0} \\
  & FedGrad~\cite{nguyen2023fedgrad}
                  & {100}   & 57.4   & {99.3}   & 57.2   & 45.5   & 80.2 \\
  & FLAME~\cite{nguyen2022flame}        & {100}   & 31.9   &  {100}   &  31.9   &  {100}  &  32.0 \\
  & FLShield~\cite{kabir2024flshield}      & 22.8   & {1.8}   & 25.8   & {1.4}   & 6.2   & {6.6} \\
  & Snowball$\boxminus$~\cite{qin2024resisting}      & {100}   & 87.6   & {100}  & 87.6   & {100}   & 87.6 \\
  & AlignIns~\cite{xu2025detecting}      & 35.7   & 49.2   &  81.2   &  41.2   &  99.3   &  27.8 \\
  & SPMC~\cite{he2025spmc}          & 1.0   & 7.2   & 1.0   & 7.2   & 1.0   & 7.2 \\
\cmidrule(lr){2-8}
  & \textbf{FedSurrogate}
                  & \cellcolor{lightgray} \textbf{100} & \cellcolor{lightgray} \textbf{1.1} & \cellcolor{lightgray} \textbf{100} & \cellcolor{lightgray} \textbf{1.1}
                  & \cellcolor{lightgray} \textbf{99.5} & \cellcolor{lightgray} \textbf{1.2} \\

\midrule

\multirow{9}{*}{\rotatebox[origin=c]{90}{\textbf{F-MNIST}}}
  & FoolsGold~\cite{fung2018mitigating}\textsuperscript{$\dagger$}
                  & 34.7   & 27.3   & 28.7   & 27.6   & 32.5   & 26.9 \\
  & FedGrad~\cite{nguyen2023fedgrad}
                  &  {100}   &  48.4   & {100}   & 48.7   &  {100}   &  48.4 \\
  & FLAME~\cite{nguyen2022flame}         &  {100}   &  31.5   &  {99.8}   & {31.5}   &  {100}  & {31.4} \\
  & FLShield~\cite{kabir2024flshield}      & \cellcolor{lightgray} \underline{99.0}   & \cellcolor{lightgray} \textbf{1.0}   & 25.5   & {1.0}   & 6.8   & {5.8} \\
  & Snowball$\boxminus$~\cite{qin2024resisting}      & {100}   & 87.6   & {99.8}  & 87.7   & {100}   & 87.6 \\
  & AlignIns~\cite{xu2025detecting}      & 29.7   & 57.5   & 37.5   & 58.0   & {32.8}   & 56.2 \\
  & SPMC~\cite{he2025spmc}          & 1.0   & 7.2   & 1.0   & 7.2   & 1.0   & 7.2 \\
\cmidrule(lr){2-8}
  & \textbf{FedSurrogate}
                  & \cellcolor{lightgray} \textbf{100} & \cellcolor{lightgray} \underline{1.1} & \cellcolor{lightgray} \textbf{99.8} & \cellcolor{lightgray}  \textbf{1.2}
                  & \cellcolor{lightgray} \textbf{100} & \cellcolor{lightgray} \textbf{1.2} \\
\midrule
\multirow{9}{*}{\rotatebox[origin=c]{90}{\textbf{CIFAR-10}}}
  & FoolsGold~\cite{fung2018mitigating}\textsuperscript{$\dagger$}
                  &  {62.0}   &  42.1   &  58.0   &  39.9   & {99.3}   &  32.5 \\
  & FedGrad~\cite{nguyen2023fedgrad}
                  & {100}   & 71.5   & {100}   & 54.2   & 1.0   & 98.0 \\
  & FLAME~\cite{nguyen2022flame}         &  {99.8}    &  {31.9}  & {99.8}   & {31.9}   & {100}   & {31.9} \\
  & FLShield~\cite{kabir2024flshield}      & 12.5   & 9.3   & 12.5   & {5.4}   & 12.0   & {6.5} \\
  & Snowball$\boxminus$~\cite{qin2024resisting}      & {100}   & 87.6   & {100}   & 87.6   & {100}   & 87.6 \\
  & AlignIns~\cite{xu2025detecting}      &   33.5 &  40.3   & 35.0   & 42.2   & 30.5   & 41.9 \\
  & SPMC~\cite{he2025spmc}  & 1.0   & {7.2}   & 1.0   & {7.2}   & 1.0   & {7.2} \\
\cmidrule(lr){2-8}
  & \textbf{FedSurrogate}
                  & \cellcolor{lightgray} \textbf{99.8} & \cellcolor{lightgray} \textbf{9.1} & \cellcolor{lightgray} \textbf{99.5} & \cellcolor{lightgray} \textbf{5.2}
                  & \cellcolor{lightgray} \textbf{99.0} & \cellcolor{lightgray} \textbf{8.3} \\

\midrule

\multirow{9}{*}{\rotatebox[origin=c]{90}{\textbf{CIFAR-100}}}
  & FoolsGold~\cite{fung2018mitigating}\textsuperscript{$\dagger$}
                  & 84.0   & 21.8   &   77.0   &  10.6   & \cellcolor{lightgray} \underline{99.8}   & \cellcolor{lightgray}  \textbf{1.0} \\
  & FedGrad~\cite{nguyen2023fedgrad}
                  &  {100}   &  44.1   &  {100}   &  43.1   & 1.0   & 100 \\
  & FLAME~\cite{nguyen2022flame}         & {100}   &  31.9   &  {100}   &  31.9   & {100}   & 31.9 \\
  & FLShield~\cite{kabir2024flshield}      & 28.7   & {1.6}   & 33.3   & 5.9   & 25.8   & {1.1} \\
  & Snowball$\boxminus$~\cite{qin2024resisting}      & {100}   & 87.6   & {100}   & 87.6   & {100}   & 87.6 \\
  & AlignIns~\cite{xu2025detecting}      & \cellcolor{lightgray} \textbf{100}   & \cellcolor{lightgray} \underline{6.2}   & \cellcolor{lightgray} \textbf{100}   & \cellcolor{lightgray} \underline{9.7}   &  95.8   &  28.9 \\
  & SPMC~\cite{he2025spmc}          & 1.0   & {1.0}   & 1.0   & {1.0}   & 1.0   & {1.0} \\
\cmidrule(lr){2-8}
  & \textbf{FedSurrogate}
                  & \cellcolor{lightgray} \underline{99.3} & \cellcolor{lightgray} \textbf{5.9} & \cellcolor{lightgray} \underline{99.8} & \cellcolor{lightgray} \textbf{2.3}
                  & \cellcolor{lightgray} {98.8} & \cellcolor{lightgray} \underline{7.9} \\

\bottomrule
\end{tabular}

\vspace{2pt}
{\footnotesize
\textsuperscript{$\dagger$}Weight-based: clients with weight ${<}\,0.5$ classified as detected.
}
\end{table}

\begin{figure}[t]
  \centering
  \includegraphics[width=0.65\textwidth]{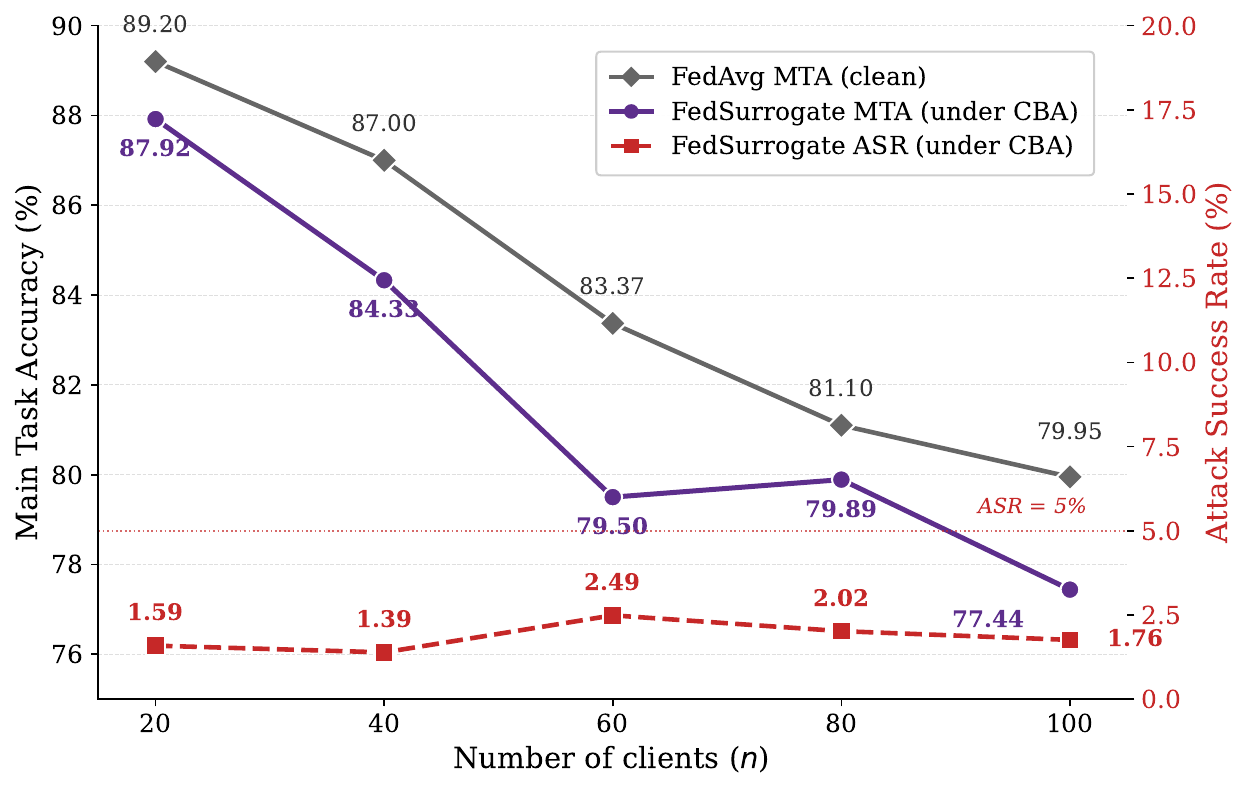}
  \caption{Scalability of \textit{FedSurrogate} on CIFAR-10 with ResNet-18 as the number of clients $n$ increases from 20 to 100, holding MCR\,$=0.2$ and $pdr=0.3$. \textit{FedSurrogate} ASR (right axis) remains below 2.5\% across the full sweep, while \textit{FedSurrogate} MTA (left axis) tracks the attack-free FedAvg baseline closely, with the gap fluctuating between 1.21 and 3.87 percentage points and showing no upward trend. The MTA degradation observed in both curves reflects the intrinsic difficulty of federated training under finer non-IID partitioning rather than over-filtering by the defense. The dotted line at ASR\,$=5\%$ marks a conventional defense-success threshold for reference.}
  \label{fig:client_scalability}
\end{figure}

\textbf{Evaluation with larger federations.}
\label{sec:scalability}
We additionally evaluate \textit{FedSurrogate} on larger client populations to characterise how its detection capability and utility cost scale with the size of the federation. We sweep $n \in \{20, 40, 60, 80, 100\}$ on CIFAR-10 with ResNet-18 under the centralised backdoor attack, holding MCR\,$=0.2$ and $pdr=0.3$ fixed, and additionally run an attack-free FedAvg baseline at each client count to isolate the utility cost of the defense from the intrinsic effects of finer partitioning.
As shown in Fig.~\ref{fig:client_scalability}, ASR remains tightly bounded between 1.39\% and 2.49\% across the full sweep with no monotonic trend, indicating that the detection capability of \textit{FedSurrogate} does not degrade as the client population grows. The MTA of \textit{FedSurrogate} declines from 87.92\% at $n=20$ to 77.44\% at $n=100$, but the attack-free FedAvg baseline exhibits a comparable decline from 89.20\% to 79.95\% over the same range, confirming that the loss reflects FL convergence under finer non-IID partitioning rather than defense over-filtering~\cite{li2020federated,zhao2018federated}. The gap between \textit{FedSurrogate} and undefended FedAvg fluctuates between 1.21 and 3.87 percentage points with no upward trend, showing that the residual utility cost of the defense remains bounded and does not scale with the client population.

\subsubsection{Resilience to Adaptive and Layer-Aware Attacks.}
\label{sec:adaptive_attacks}
We evaluate \textit{FedSurrogate} under a stronger adversarial setting where the attacker has full knowledge of the defense pipeline. We consider three attacks: the \emph{LP attack}~\cite{zhuang2024backdoor}, which identifies backdoor-critical layers via forward and backward layer substitution and poisons only those layers while keeping the rest benign; and two defense-aware attacks that we design. The \emph{Adaptive Cosine Similarity Attack (CSA)} assumes each malicious client first trains a benign reference model on clean data, then disguises the backdoored model by optimising a layer-wise cosine similarity loss between the backdoor update and the benign update. The \emph{Adaptive Critical Layer Attack (CLA)} extends CSA by selecting the top-$k$ layers with the highest cosine similarity and replacing their benign parameters with backdoored ones, while keeping the remaining layers benign. We set the cosine similarity loss weight to $\lambda=1$ for CNN and $\lambda=5$ for ResNet-18, searched from $\{0.5,1,2,5,10\}$, and $k=2$ for CNN and $k=4$ for ResNet-18.

\begin{table}[!htbp]
\centering
\caption{\emph{FedSurrogate} against adaptive and layer-aware attacks (MCR\,$=0.2$, $pdr=0.3$).}
\label{tab:adaptive_attacks}
\setlength{\tabcolsep}{4pt}
\renewcommand{\arraystretch}{1.1}
\small
\begin{tabular}{@{}ll|cccc@{}}
\toprule
\textbf{Dataset} & \textbf{Attack} & MTA$\uparrow$ & ASR$\downarrow$ & TPR$\uparrow$ & FPR$\downarrow$ \\
\midrule
\multirow{3}{*}{\textbf{CIFAR-10}}
  & LP~\cite{zhuang2024backdoor} & 84.69 & 1.08 & 95.5 & 16.4 \\
  & CSA & 87.54 & 1.28 & 99.8 & 9.9 \\
  & CLA & 87.59 & 2.54 & 99.8 & 3.1 \\
\midrule
\multirow{3}{*}{\textbf{CIFAR-100}}
  & LP~\cite{zhuang2024backdoor} & 63.81 & 0.18 & 99.8 & 1.8 \\
  & CSA & 66.80 & 0.26 & 100 & 2.8 \\
  & CLA & 66.51 & 0.22 & 100 & 1.2 \\
\bottomrule
\end{tabular}
\end{table}

As shown in Table~\ref{tab:adaptive_attacks}, \textit{FedSurrogate} maintains ASR below 2.6\% against all three attacks on both datasets. The LP attack causes the most noticeable impact on CIFAR-10, with MTA dropping to 84.69\% and FPR rising to 16.4\%, as its minimal-layer poisoning creates a harder separation problem for density-based clustering; on CIFAR-100, LP has minimal effect. The defense-aware CSA and CLA attacks are effectively neutralised: the cosine-alignment regularisation makes the malicious update directionally similar to benign updates, but the temporal score aggregation in Stage~2 still accumulates sufficient evidence across rounds to distinguish persistent adversaries from transient heterogeneity. Notably, CLA achieves the lowest FPR across both datasets (3.1\% and 1.2\%) because its non-selected layers are genuinely benign, but this evasion strategy simultaneously limits the attack's effectiveness as reflected in the low ASR.

\section{Conclusion}
We presented \textit{FedSurrogate}, a server-side defense against backdoor attacks in FL. \textit{FedSurrogate} integrates a Layer Criticality Analysis mechanism to identify, in each round, a small subset of model layers carrying the strongest backdoor signal through directional cosine divergence and restricts both clustering and gradient-alignment scoring to that subspace.
Building on LCA, we design a bidirectional alignment filter that rescues benign clients erroneously flagged under non-IID heterogeneity, and a surrogate-replacement mechanism that substitutes only the critical layers of confirmed malicious updates with downscaled benign counterparts, preserving main-task knowledge while neutralising the backdoor. Across four benchmarks, two architectures, and three attack families including the stealth-optimised Neurotoxin attack, \textit{FedSurrogate} consistently ranks among the top methods in utility and security, achieves the lowest FPR across all settings, and remains robust to larger client populations and operating points.

\clearpage

\clearpage

\bibliographystyle{splncs04}
\bibliography{references}

\clearpage

\section*{Appendix}

\subsection*{Ablation Studies}



\begin{figure}[!htbp]
  \centering
  \includegraphics[width=0.7\textwidth]{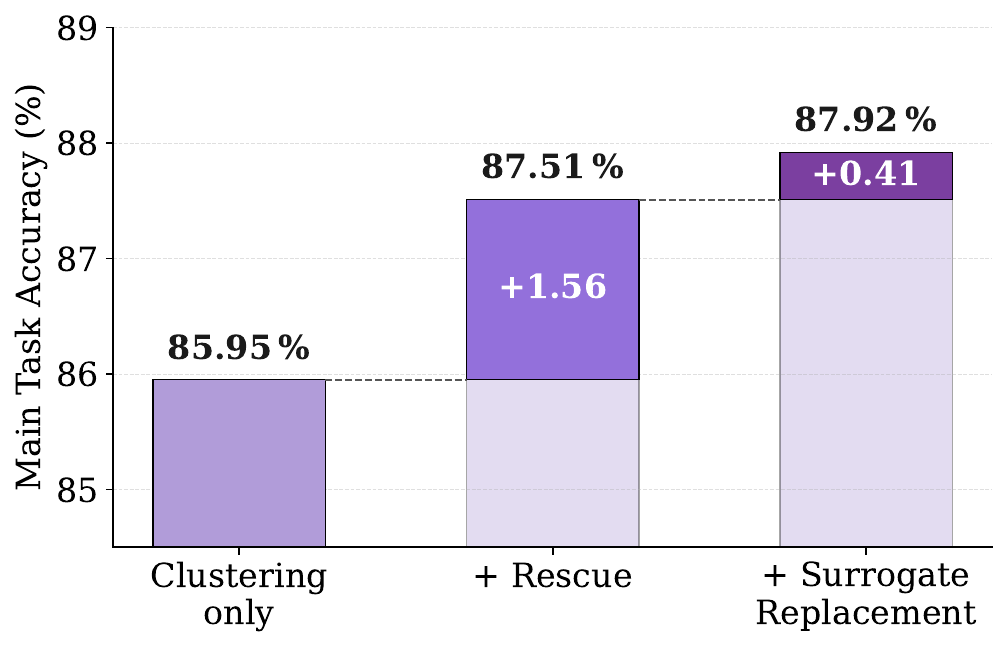}
  \caption{Ablation study of \textit{FedSurrogate} components on CIFAR-10 with centralized backdoor attack (CBA)
  Each component contributes incrementally to MTA, with the rescue stage
  providing the largest single gain (+1.56\%).}
  \label{fig:ablation_cifar10}
\end{figure}

 \vspace{-15pt}


\subsection*{Effect of Donor Selection Strategy}

\vspace{-30pt}


\begin{table}[!htbp]
\centering
\caption{Effect of the donor selection metric in the surrogate
replacement stage of \textit{FedSurrogate} on CIFAR-10
(CBA, $pdr{=}0.3$, $n{=}20$). Cosine similarity is the default in
our implementation.}
\label{tab:donor_metric}
\setlength{\tabcolsep}{8pt}
\renewcommand{\arraystretch}{1.15}
\small
\begin{tabular}{@{}l|cc@{}}
\toprule
\textbf{Donor Metric} & MTA$\uparrow$ & ASR$\downarrow$ \\
\midrule
Euclidean distance & 85.86 & 1.56 \\
\textbf{Cosine similarity (ours)} & \textbf{87.92} & 1.59 \\
\bottomrule
\end{tabular}
\end{table}


\noindent\textbf{Donor selection metric.} Table~\ref{tab:donor_metric} compares metrics for Stage~3 donor selection. Cosine similarity yields a 2.06pp MTA improvement over Euclidean distance at identical ASR, as it matches donors by directional alignment rather than magnitude.

\begin{figure}[h]
  \centering
  \includegraphics[width=0.7\textwidth]{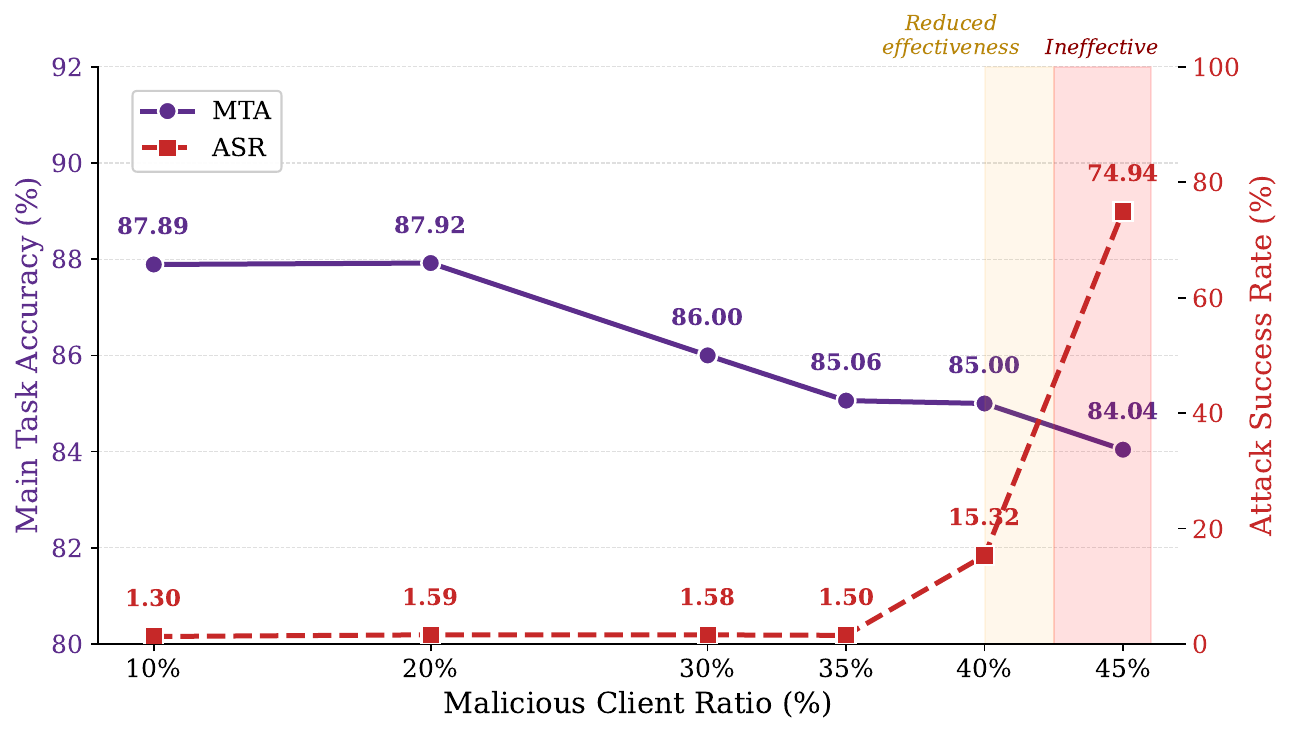}
  \caption{Robustness of \textit{FedSurrogate} against increasing
Malicious Client Ratio (MCR) on CIFAR-10 under the CBA attack
(Dirichlet $\alpha{=}0.5$, $n{=}20$ clients). The defense maintains
ASR below 2\% for MCR $\leq 35\%$, enters a reduced-effectiveness
regime at MCR $=$ 40\% (ASR $=$ 15.32\%) where the attack partially
succeeds but remains substantially mitigated relative to the
undefended case, and becomes ineffective at MCR $=$ 45\%
(ASR $=$ 74.94\%)}
\label{fig:mcr_sensitivity_cifar10}
\end{figure}

\noindent\textbf{Sensitivity to malicious client ratio.} Fig.~\ref{fig:mcr_sensitivity_cifar10} reports MTA and ASR as MCR increases from 0.10 to 0.45 on CIFAR-10 (ResNet-18, CBA, $pdr=0.3$). FedSurrogate maintains ASR below 1.6\% for MCR\,$\leq 0.35$, with MTA degrading mildly from 87.89\% to 85.06\%. Beyond this point, ASR rises to 15.32\% at MCR\,$=0.40$ and 74.94\% at MCR\,$=0.45$, while MTA remains stable at 84--85\%, indicating that detection failures manifest as missed adversaries rather than false rejections. We characterise FedSurrogate as effective for MCR\,$\leq 0.35$, well above the standard setting of MCR\,$=0.2$.


\begin{table*}[b]
\centering
\caption{MTA and ASR results of baselines and \textit{FedSurrogate} on IID CIFAR-10 and CIFAR-100.
Results are shown in \%. The best result in each column is \textbf{bold} and the second best is \underline{underlined}, considering only defenses that achieve both: MTA within $5\%$ of FedAvg's MTA, and successfully mitigate the attacks (ASR $< 5\%$), marked in \colorbox{lightgray}{light grey}.}
\label{tab:main_results_cifar_updated}
\scriptsize
\resizebox{\textwidth}{!}{%
\renewcommand{\arraystretch}{1.15}
\begin{tabular}{@{}ll|cc|cc|cc|cc@{}}
\toprule
\multirow{2}{*}{\textbf{Dataset}}
  & \multirow{2}{*}{\textbf{Methods}}
  & \multicolumn{2}{c|}{\textbf{CBA}}
  & \multicolumn{2}{c|}{\textbf{DBA}}
  & \multicolumn{2}{c|}{\textbf{Neurotoxin}}
  & \multirow{2}{*}{\textbf{Avg. MTA$\uparrow$}}
  & \multirow{2}{*}{\textbf{Avg. ASR$\downarrow$}} \\
\cmidrule(lr){3-4}\cmidrule(lr){5-6}\cmidrule(lr){7-8}
  &
  & MTA$\uparrow$ & ASR$\downarrow$
  & MTA$\uparrow$ & ASR$\downarrow$
  & MTA$\uparrow$ & ASR$\downarrow$
  & & \\
\midrule

\multirow{10}{*}{\rotatebox[origin=c]{90}{\textbf{CIFAR-10}}}
  & FedAvg
  & 92.04 & 99.34
  & 92.27 & 21.21
  & 92.16 & 97.62
  & 92.16 & 72.72 \\
\cmidrule(lr){2-10}
  & FoolsGold~\cite{fung2018mitigating}
  & \cellcolor{lightgray} \underline{91.22} & \cellcolor{lightgray} \underline{0.28}
  & \cellcolor{lightgray} \underline{90.96} &\cellcolor{lightgray} 0.37
  &\cellcolor{lightgray} \underline{90.88} & \cellcolor{lightgray}0.74
  & \cellcolor{lightgray} \underline{91.02} & \cellcolor{lightgray}0.46 \\
  & FedGrad~\cite{nguyen2023fedgrad}
  & \cellcolor{lightgray}89.10 & \cellcolor{lightgray}0.60
  & \cellcolor{lightgray} 90.21 & \cellcolor{lightgray}0.43
  & 85.06 & 99.92
  & 88.12 & 33.65 \\
  & FLAME~\cite{nguyen2022flame}
  & \cellcolor{lightgray}88.88 & \cellcolor{lightgray}0.76
  & \cellcolor{lightgray}88.83 & \cellcolor{lightgray}0.70
  & \cellcolor{lightgray}88.06 & \cellcolor{lightgray}0.66
  & \cellcolor{lightgray}88.59 & \cellcolor{lightgray}0.71 \\
  & FLShield~\cite{kabir2024flshield}
  & 91.97 & 96.97
  & 91.68 & 34.56
  & 91.48 & 95.43
  & 91.71 & 75.65 \\
  & Snowball$\boxminus$~\cite{qin2024resisting}
  & \cellcolor{lightgray}89.17 &\cellcolor{lightgray} 0.40
  & \cellcolor{lightgray}87.67 & \cellcolor{lightgray}0.50
  & 84.59 & 0.76
  & 87.14 & 0.55 \\
  & AlignIns~\cite{xu2025detecting}
  & \cellcolor{lightgray} \textbf{91.42} & \cellcolor{lightgray} \textbf{0.23}
      & \cellcolor{lightgray}90.65 & \cellcolor{lightgray} \textbf{0.27}
  & \cellcolor{lightgray}90.83 & \cellcolor{lightgray} \textbf{0.29}
  & \cellcolor{lightgray}90.97 & \cellcolor{lightgray} \textbf{0.26} \\
  & SPMC~\cite{he2025spmc}
  & 91.77 & 99.09
  & 92.07 & 31.84
  & 91.80 & 98.25
  & 91.88 & 76.39 \\
\cmidrule(lr){2-10}
  & \textbf{FedSurrogate}
  & \cellcolor{lightgray}91.17 &\cellcolor{lightgray} \underline{0.28}
  & \cellcolor{lightgray} \textbf{91.15} & \cellcolor{lightgray} \underline{0.34}
  & \cellcolor{lightgray} \textbf{91.35} & \cellcolor{lightgray} \underline{0.48}
  & \cellcolor{lightgray} \textbf{91.22} & \cellcolor{lightgray} \underline{0.37} \\

\midrule

\multirow{10}{*}{\rotatebox[origin=c]{90}{\textbf{CIFAR-100}}}
  & FedAvg
  & 71.21 & 98.19
  & 71.28 & 26.68
  & 70.84 & 93.54
  & 71.11 & 72.80 \\
\cmidrule(lr){2-10}
  & FoolsGold~\cite{fung2018mitigating}
  & \cellcolor{lightgray}\textbf{69.91} & \cellcolor{lightgray} \textbf{0.12}
  & 68.89 & 6.59
  & \cellcolor{lightgray} \underline{69.05} & \cellcolor{lightgray}0.16
  & \cellcolor{lightgray} \textbf{69.28} & \cellcolor{lightgray}2.29 \\
  & FedGrad~\cite{nguyen2023fedgrad}
  & 63.38 & 0.23
  & 63.87 & 0.21
  & 48.47 & 99.98
  & 58.57 & 33.47 \\
  & FLAME~\cite{nguyen2022flame}
  & 59.66 & 0.33
  & 59.26 & 0.30
  & 59.89 & 0.13
  & 59.60 & 0.25 \\
  & FLShield~\cite{kabir2024flshield}
  & 70.78 & 93.81
  & 71.14 & 26.83
  & 70.74 & 88.95
  & 70.89 & 69.86 \\
  & Snowball$\boxminus$~\cite{qin2024resisting}
  & 56.68 & 0.20
  & 48.80 & 0.13
  & 54.83 & 0.06
  & 53.44 & 0.13 \\
  & AlignIns~\cite{xu2025detecting}
  & \cellcolor{lightgray} \underline{68.97} & \cellcolor{lightgray}\underline{0.18}
  & \cellcolor{lightgray} \underline{69.25} & \cellcolor{lightgray} \underline{0.18}
  & \cellcolor{lightgray}68.06 & \cellcolor{lightgray} \textbf{0.09}
  & \cellcolor{lightgray}68.76 & \cellcolor{lightgray} \textbf{0.15} \\
  & SPMC~\cite{he2025spmc}
  & 71.87 & 96.07
  & 71.60 & 12.91
  & 71.20 & 93.97
  & 71.56 & 67.65 \\
\cmidrule(lr){2-10}
  & \textbf{FedSurrogate}
  & \cellcolor{lightgray} 68.92 & \cellcolor{lightgray} 0.23
  & \cellcolor{lightgray} \textbf{69.34} & \cellcolor{lightgray} \textbf{0.17}
  & \cellcolor{lightgray} \textbf{69.18} & \cellcolor{lightgray} \underline{0.14}
  & \cellcolor{lightgray} \underline{69.15} & \cellcolor{lightgray} \underline{0.18} \\

\bottomrule
\end{tabular}%
}

\end{table*}



\end{document}